\documentclass[journal]{IEEEtran}

\usepackage{multirow}  
\usepackage{bm}        
\usepackage{cite}
\usepackage{array}
\usepackage{color}
\usepackage{float}
\usepackage[cmex10]{amsmath}
\usepackage{nomencl}	
\usepackage{stfloats}

\usepackage[normalem]{ulem}			
\usepackage{diagbox}						
\usepackage{slashbox}						
\usepackage{colortbl}						
\usepackage{multirow}						
\usepackage{tabularx}
\usepackage{placeins}
\usepackage{algorithm}
\usepackage[noend]{algpseudocode}
\usepackage{siunitx}
\usepackage{gensymb}

\ifCLASSINFOpdf
  \usepackage[pdftex]{graphicx}
	\graphicspath{{./figure/}}
  \DeclareGraphicsExtensions{.pdf,.jpeg,.png}
\else
  \usepackage[dvips]{graphicx}
  \graphicspath{{./figure/}}
  \DeclareGraphicsExtensions{.eps}
\fi

\ifCLASSOPTIONcompsoc
  \usepackage[caption=false,font=normalsize,labelfont=sf,textfont=sf]{subfig}
\else
  \usepackage[caption=false,font=footnotesize]{subfig}
\fi

\newcommand{\figref}[1]{\textcolor{black}{Fig.~\ref{#1}}}
\newcommand{\tableref}[1]{\textcolor{black}{Table~\ref{#1}}}

\numberwithin{equation}{section}
\renewcommand{\theequation}{\arabic{section}.\arabic{equation}}

\counterwithout{equation}{section}
\begin{document}
\bstctlcite{IEEEexample:BSTcontrol}
\setcounter{page}{1}

\newpage

\onecolumn
\vspace*{0pt}
This work will be submitted to the IEEE for possible publication. Copyright may be transferred without notice, after which this version may no longer be accessible.
\newpage
\twocolumn

\title{Hybrid Voltage-Current Control of Grid-Forming and Grid-Following Inverters}
\author{Zirui Wang, Yitong Li, \IEEEmembership{Member, IEEE} Quanchi Wu,  Jinjun Liu, \IEEEmembership{Fellow, IEEE}}

\ifCLASSOPTIONpeerreview
	\maketitle 
\else
	\maketitle
\fi


\begin{abstract}
Grid-connected inverters are required to operate stably under a wide range of grid conditions. However, conventional grid-following (GFL) control may suffer from instability under weak-grid conditions, while grid-forming (GFM) control may exhibit unstable oscillations under strong-grid conditions. To address these issues, a hybrid voltage–current control method is proposed in this article. A voltage control is introduced on the $d$-axis, while a current control is adopted on the $q$-axis, enabling the inverter to exhibit voltage-source characteristics on the $d$-axis and current-source characteristics on the $q$-axis. In this way, the proposed control integrates the characteristics of both conventional GFL and GFM control. A full-order model is established to analyze the port characteristics and small-signal stability of the systems. Finally, the effectiveness of the proposed control strategy is validated through simulations and experiments on a 1.5 kW inverter experimental platform. The results show that the proposed control maintains stable operation under different grid conditions with varying short-circuit ratios (SCRs).
\end{abstract}




\section{Introduction} \label{Section:Introduction}

With the large-scale integration of renewable energy and distributed generation, the proportion of inverter-based resources in power systems continues to rise. As the penetration of synchronous generators decreases, system inertia and voltage/frequency support are weakened. This makes stability issues more likely to occur, especially in weak grids or during large disturbances\cite{rocabert2012control}. In this context, stable operation of inverters under varying grid strengths has become a key issue for both academia and industry. Mainstream control strategies of inverters can be divided into grid-following (GFL) and grid-forming (GFM)\cite{blaabjerg2006overview}. A GFL inverter serves as a current source and uses a phase-locked loop (PLL) to synchronize with the grid voltage phase. In contrast, a GFM inverter mimics the synchronous generator characteristics and operates as a voltage source to enhance voltage and frequency support\cite{zhong2010synchronverters},\cite{zhang2009power}.

However, these two single-mode control approaches have inherent limitations under complex grid conditions. Each approach performs well only within a certain range of short-circuit ratio (SCR)\cite{yang2014impedance},\cite{sun2011impedance}. GFL control relies on a stable grid voltage. \cite{han2023stability} and \cite{liserre2006stability} shows that, in weak grids or under large disturbances, the coupling between the PLL and the grid impedance can significantly reduce synchronization stability. By contrast, while GFM control offers higher voltage support in weak-grid scenarios, but overcurrent and instability issues may occur under strong-grid conditions\cite{matevosyan2019grid},\cite{wu2018sequence}. In conclusion, it is difficult to achieve both good dynamic performance and stability under different grid strengths and disturbances\cite{rosso2021grid}.

To combine the advantages of GFM and GFL inverters, unified and hybrid control architectures are explored in literature. \cite{geng2022unified} and \cite{harnefors2020universal} proposed unified control frameworks for both GFL and GFM inverters. These frameworks aim to adapt to different grid strengths using a single control structure. \cite{gao2024seamless} proposed seamless switching between GFL and GFM modes based on grid conditions. \cite{lima2022hybrid} combines the modulation signals of GFL and GFM using weighted blending. This allows the inverter to exhibit partial GFL and GFM characteristics in specific frequency ranges. \cite{xiao2023adaptive} and \cite{harnefors2020reference} introduce specific feedback mechanisms to realize series or parallel combinations of GFL and GFM control at the algorithm level. \cite{liu2022physical} designed an adaptive synchronization method that combines a PLL and a power-synchronization loop (PSL). It adjusts the weighting between the two loops under different disturbances and grid strengths. As a result, the inverter can incorporate both GFL and GFM features.

Although above strategies improve system performance, most studies focus on merging synchronization loops or mixing operating \cite{gong2023interaction},\cite{yu2024unified},\cite{tafizare2024grid}. In these designs, the dq axis control channels are typically treated symmetrically. Consequently, the potential of assigning both voltage and current source characteristics to the $dq$ axes is rarely explored\cite{li2020impedance},\cite{ai2024extension}. \cite{li2022revisiting} revisits GFM and GFL control using duality theory. It reveals the intrinsic correspondence between their control structures and dynamic behaviors, and provides a theoretical basis for hybrid control frameworks. Motivated by these insights, this paper proposes an voltage-current amplitude hybrid control method. The proposed architecture assigns voltage-source behavior to the $d$-axis and current-source behavior to the $q$-axis. With this structure, the inverter achieves improved stability margins across different grid strengths, while voltage support and current regulation capabilities are preserved.

The remainder of this paper is organized as follows. Section II presents the proposed hybrid voltage–current control and its $dq$-frame structure, where the inverter behaves as a controlled voltage source in the $d$-axis and a controlled current source in the $q$-axis. State-space and impedance modeling are then employed to analyze the port impedance and small-signal stability, showing that the proposed control extends the stable SCR operating range of conventional GFM and GFL control. Section III provides comparative simulation and experimental results under four control modes. Finally, Section VI concludes the paper.


\section{Hybrid Voltage-Current Control of Inverters} \label{Section:Main}

\subsection{Control Structure}\label{Hybrid Voltage-Current Control}

Based on existing studies, hybrid control structure have been mainly developed by combining different synchronization mechanisms to improve the performance of grid-connected inverters. Motivated by this, this paper further extends the concept of hybridization to the control loops themselves, by exploring the intrinsic voltage-source and current-source characteristics of the inverter dynamics. Accordingly, a hybrid voltage–current control structure is proposed, whose control structure is shown in Fig. 1. The proposed method is compatible with both $P-\omega$ droop control and PLL synchronization, as well as other synchronization schemes.

Unlike conventional GFL or GFM control, which preserve symmetry between the $d$‑axis and $q$‑axis, hybrid voltage-current control differentiates the control objectives across axes to balance voltage stiffness and current response speed. The $d$-axis adopts a dual-loop voltage–current control structure, enabling the inverter to actively regulate the port voltage magnitude and thereby exhibit certain GFM characteristics. The $d$-axis control law is expressed as:
\begin{equation}
\left\{
\begin{aligned}
& v_{id} - v_{cd} = (sL_f + R_f)i_d \\
& Z_{\text{PI}i,d}\left((K_{pv,d} + \frac{K_{iv,d}}{s})(v_{cd}^*-v_{cd}) - i_{ld}\right) = v_{id} \\
& Z_{\text{PI}i,d} = (K_{pi,d} + \frac{K_{ii,d}}{s})G_{\text{del}} \\
& i_{ld} = i_{gd} + sC_f v_{cd}
\end{aligned}
\right.
\end{equation}

\begin{figure}[!t]
\centering
\includegraphics[scale=0.5]{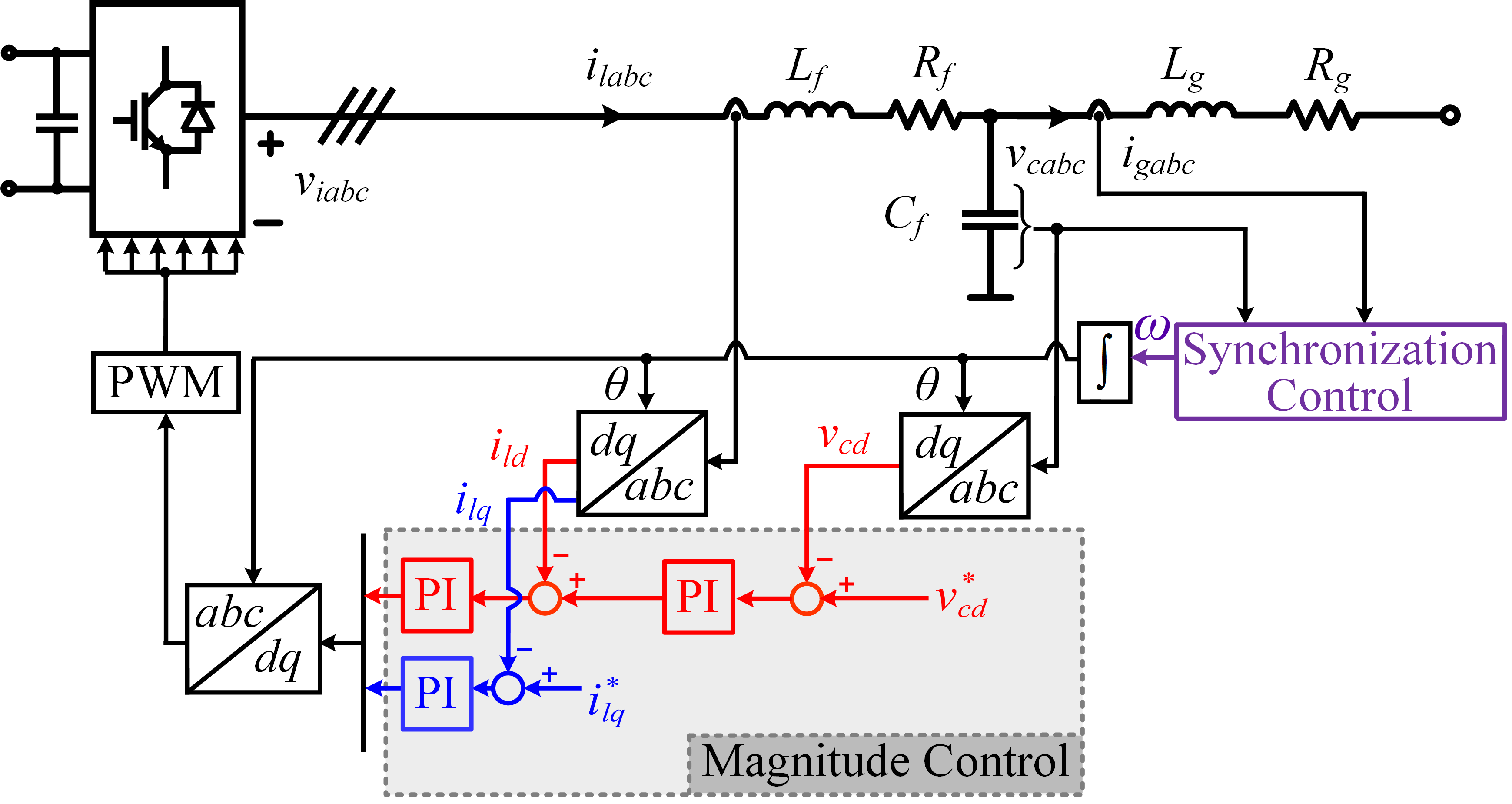}
\caption{Control structure of hybrid voltage-current inverter.}
\label{Fig:control structure}
\end{figure} 
Finally, the $d$-axis Thevenin equivalent equation can be derived.
The $q$-axis employs current-loop control, primarily used for reactive power regulation and current limitation, thereby exhibiting certain GFL characteristics.The $q$-axis control law is expressed as:
\begin{equation}
\left\{
\begin{aligned}
& v_{iq} - v_{cq} = (sL_f + R_f)i_q \\
& (K_{pi,q} + \frac{K_{ii,q}}{s})G_{\text{del}} = Z_{\text{PI}i,q} \\
& Z_{\text{PI}i,q}(i_{lq}^* - i_{lq}) = v_{iq}
\end{aligned}
\right.
\end{equation}
Finally, the $q$-axis Norton equivalent equation can be derived as:
\begin{equation}
i_{lq} = \frac{Z_{\text{PI}i,q}}{Z_{\text{PI}i,q} + sL_f + R_f} i_{lq}^*
- \frac{1}{Z_{\text{PI}i,q} + sL_f + R_f} v_{cq}
\label{eq:iq_norton}
\end{equation}
By asymmetrically configuring the voltage and current loops on the $d$‑axis and $q$‑axis, the proposed hybrid voltage–current control strategy effectively integrates grid-forming and grid-following characteristics. Specifically, the $d$‑axis voltage control channel provides the inverter with voltage magnitude regulation, while the $q$‑axis current control channel ensures current control capability. The inverter thus operates as a controlled voltage source in the $d$‑axis dynamics and as a controlled current source in the $q$‑axis dynamics, as depicted in \figref{Fig:dq behavior}. As a result, the inverter no longer falls into the conventional category of purely GFM or purely GFL, but instead represents a hybrid grid-connected inverter with a more flexible allocation of control degrees of freedom.

\subsection{Port Characteristics}\label{Voltage Forming (VFM): Much Smaller Port Impedance}

\begin{figure}[htbp]
\centering
\includegraphics[scale=0.33]{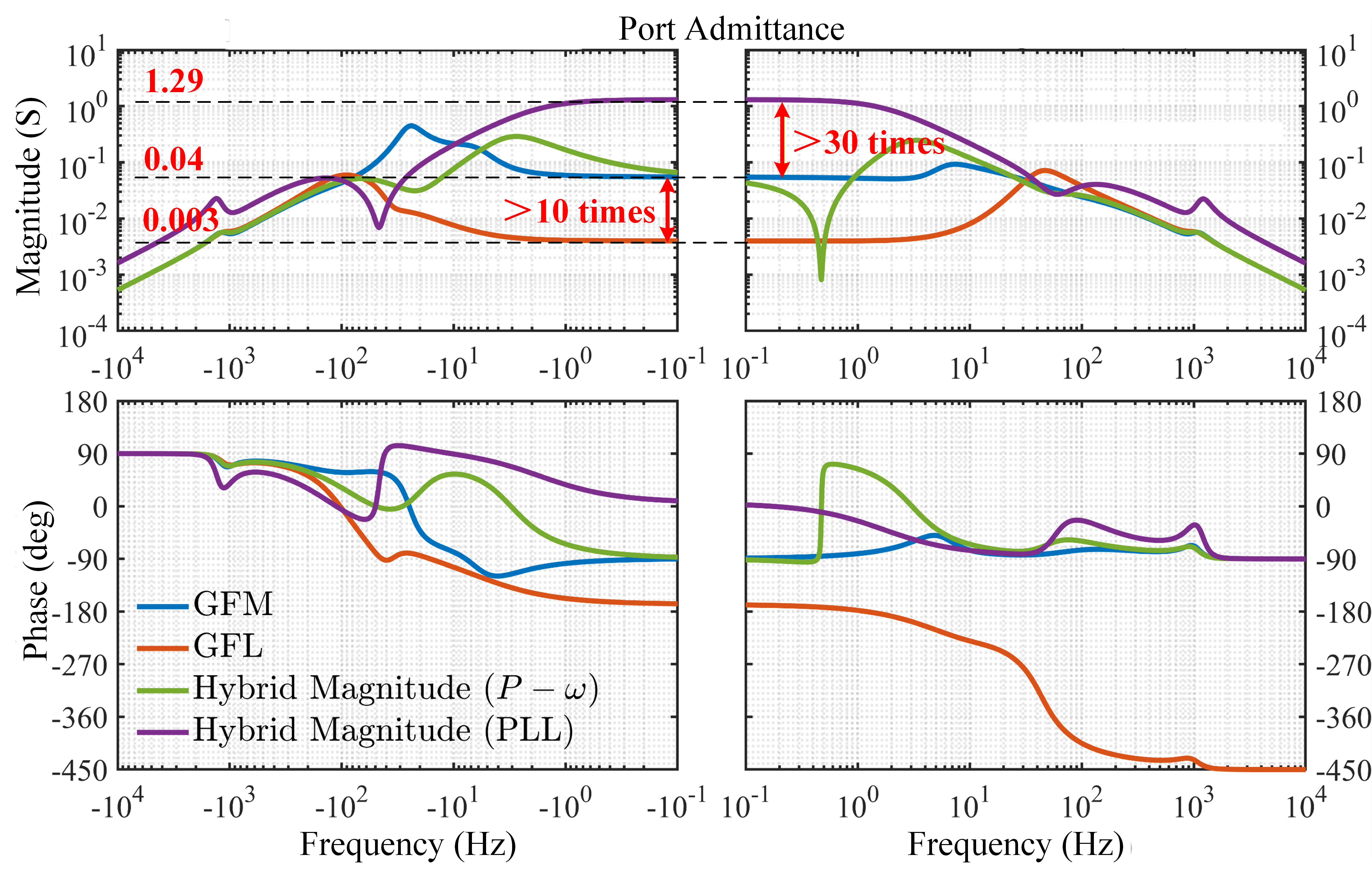}
\caption{Port admittance comparison among GFM, GFL,
hybrid voltage-current control in the complex vector dq frame.}
\label{Fig:Port admittance}
\end{figure}

\begin{figure*}[!t]
\centering
\includegraphics[scale=0.20]{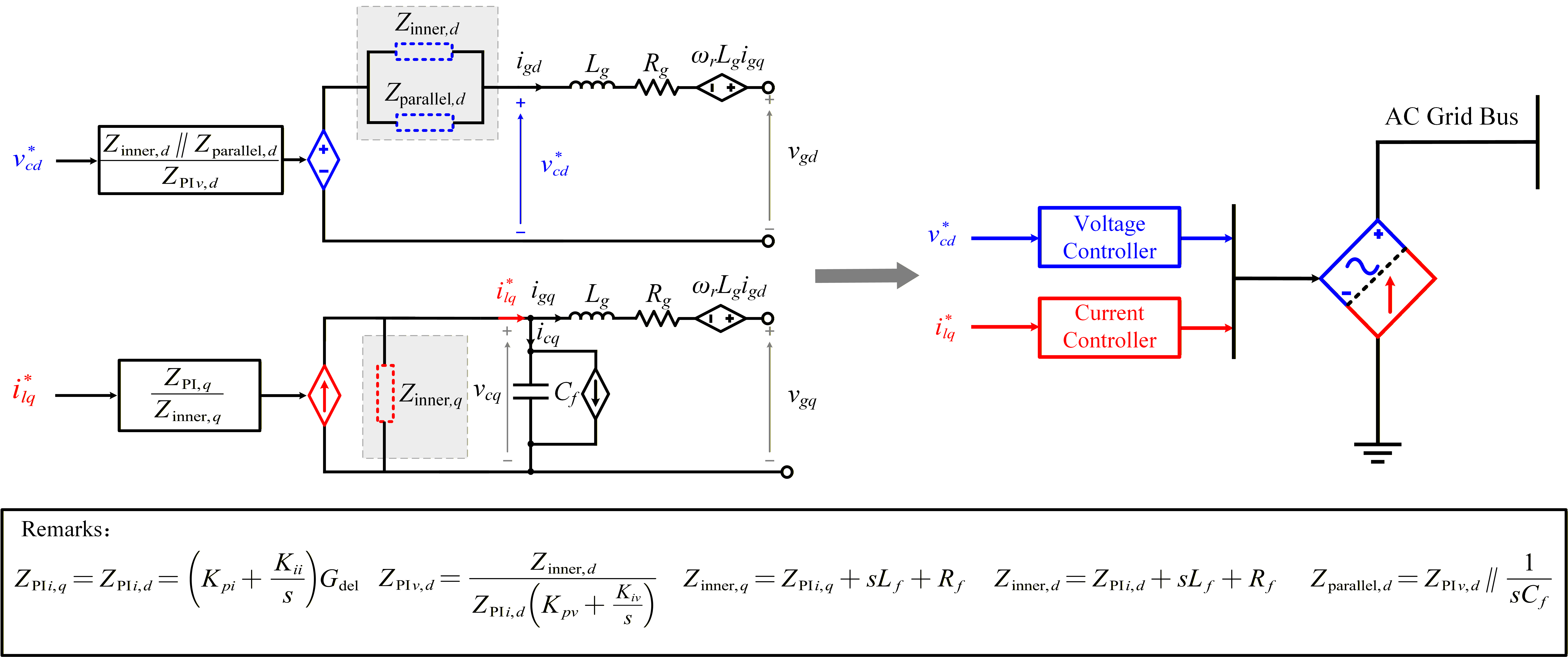}
\caption{Modeling the hybrid voltage-current control of inverter.}
\label{Fig:dq behavior}
\end{figure*}

An essential criterion for evaluating the voltage-forming capability of an inverter is the magnitude of its port impedance. Typically, a voltage source exhibits a much smaller port impedance (or equivalently a larger port admittance) than a current source. The port impedance (or equivalently, the admittance) of the four control modes can be derived based on the state-space modeling approach. The parameters of the grid-tied inverter system and its control are listed in \tableref{GRID-TIED INVERTER SYSTEM PARAMETERS} and \tableref{CONTROL PARAMETERS OF FOUR METHOD}, which are also used in the subsequent experimental setup and control implementation. 

\figref{Fig:Port admittance} compares the port admittance of four control structures, including GFM, GFL, and the proposed hybrid magnitude control under different synchronization methods. It can be clearly observed that around 0 Hz (i.e., the fundamental frequency in the natural frame), the port admittance of the hybrid magnitude control is significantly larger than that of the conventional control schemes. Specifically, in the low-frequency region, the port admittance of the GFL control is much smaller than that of the other control schemes. The voltage–current control under PLL-based synchronization not only exhibits a port admittance even larger than that of the GFM control, but also introduces positive damping characteristics. Meanwhile, the GFM control lies between the droop-based voltage–current control and the conventional GFL control. These results indicate that the proposed control strategy can effectively enhance the voltage-forming capability of the inverter and significantly reduce the port impedance.

\begin{figure*}[!t]
\centering
\subfloat[]{\includegraphics[width=3.5in]{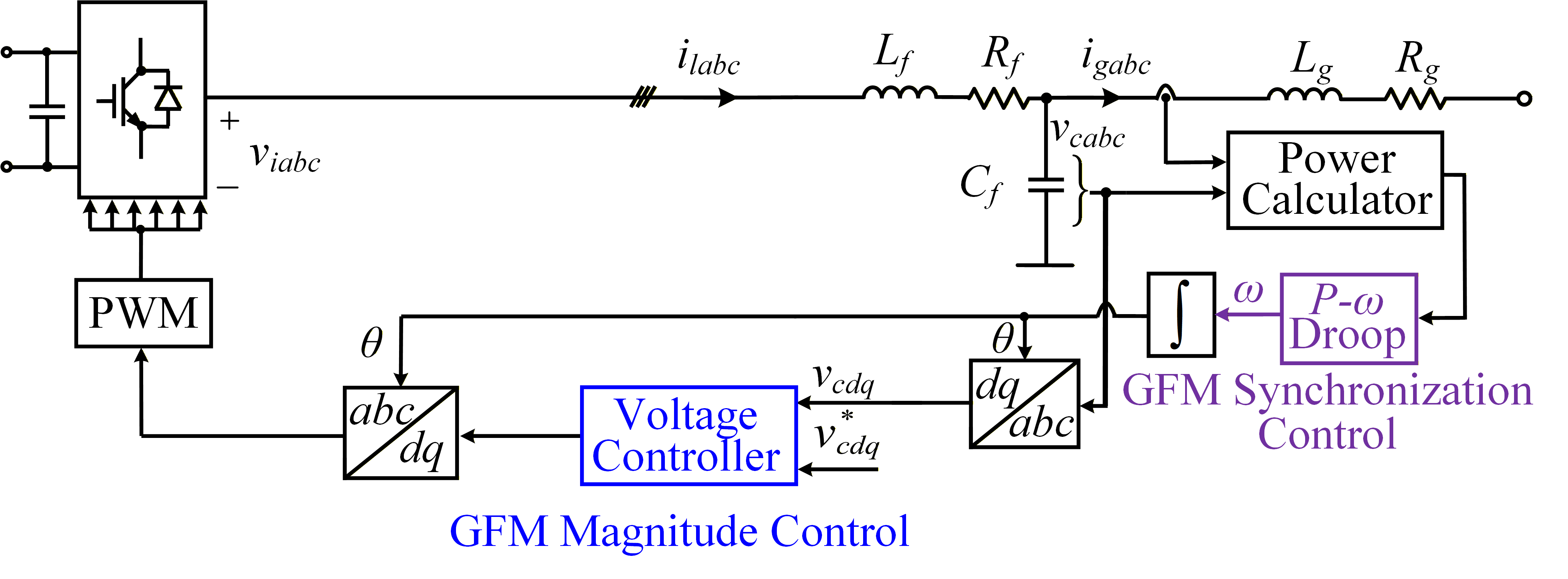}%
\label{fig:GFM}}
\hfil
\subfloat[]{\includegraphics[width=3.5in]{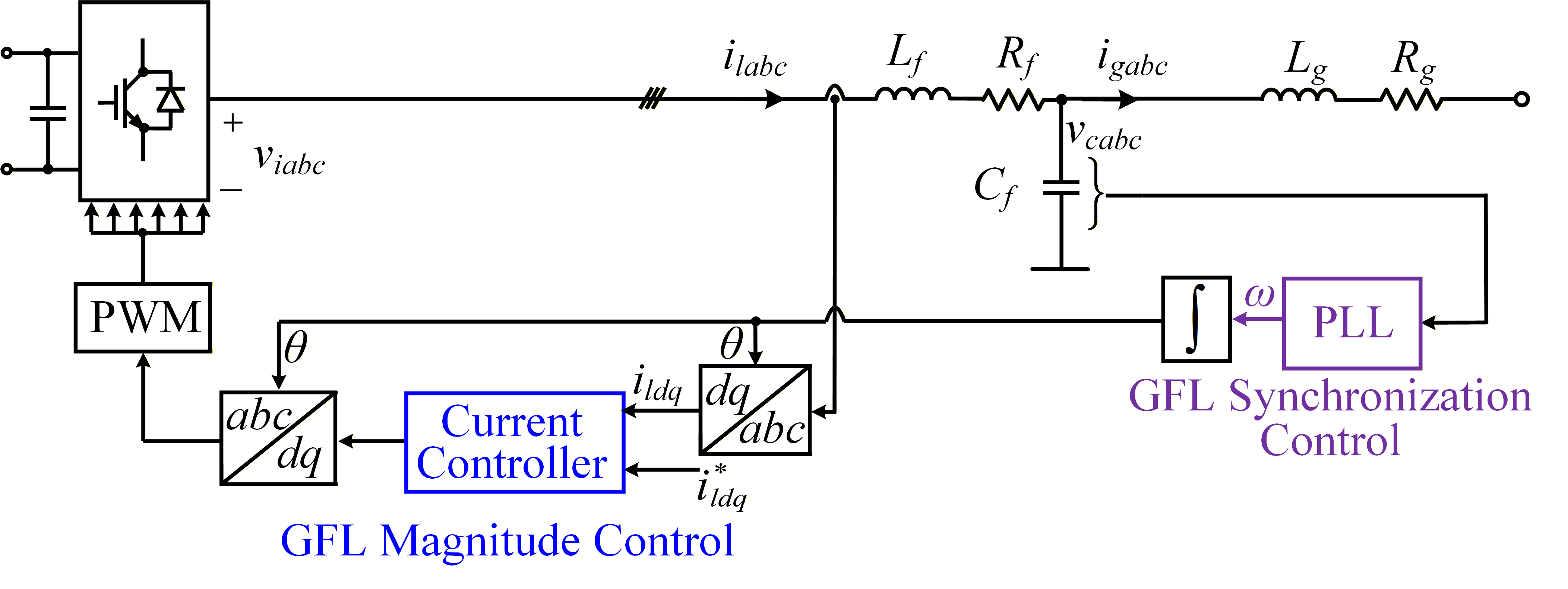}%
\label{fig:GFL}}
\hfil
\subfloat[]{\includegraphics[width=3.5in]{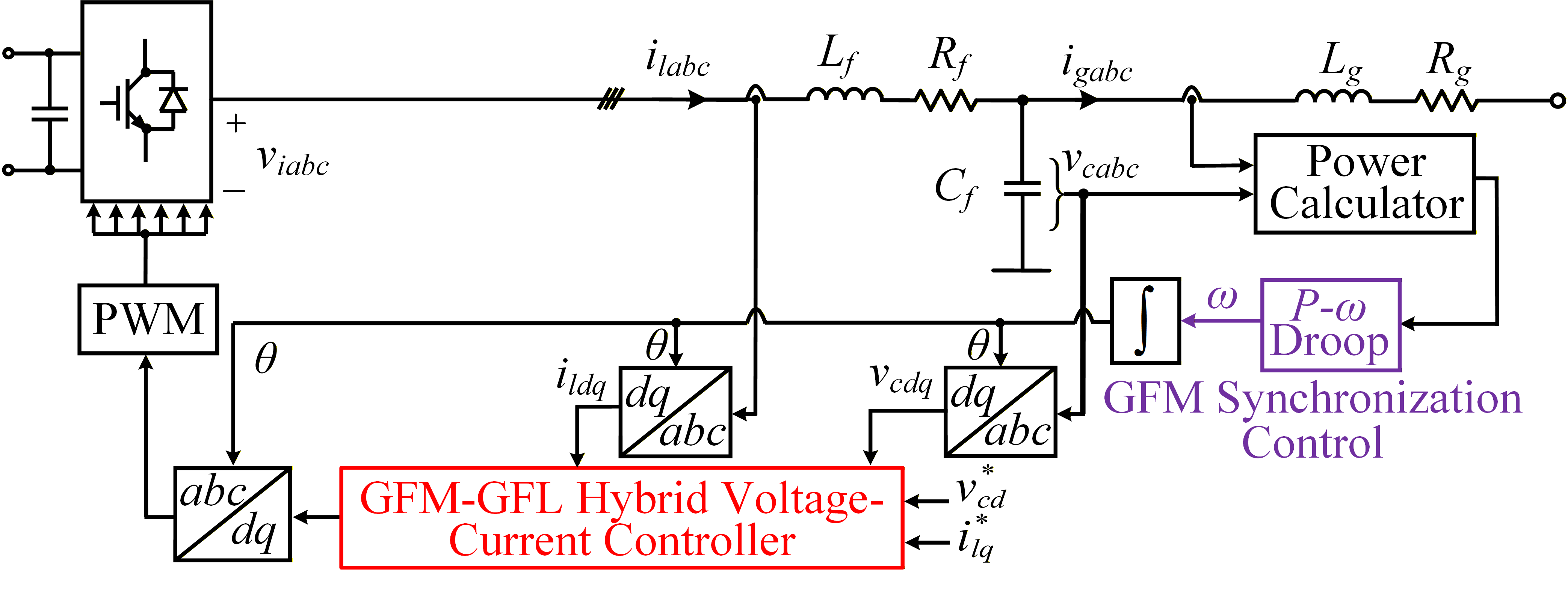}%
\label{fig:vdiqdroop}}
\hfil
\subfloat[]{\includegraphics[width=3.5in]{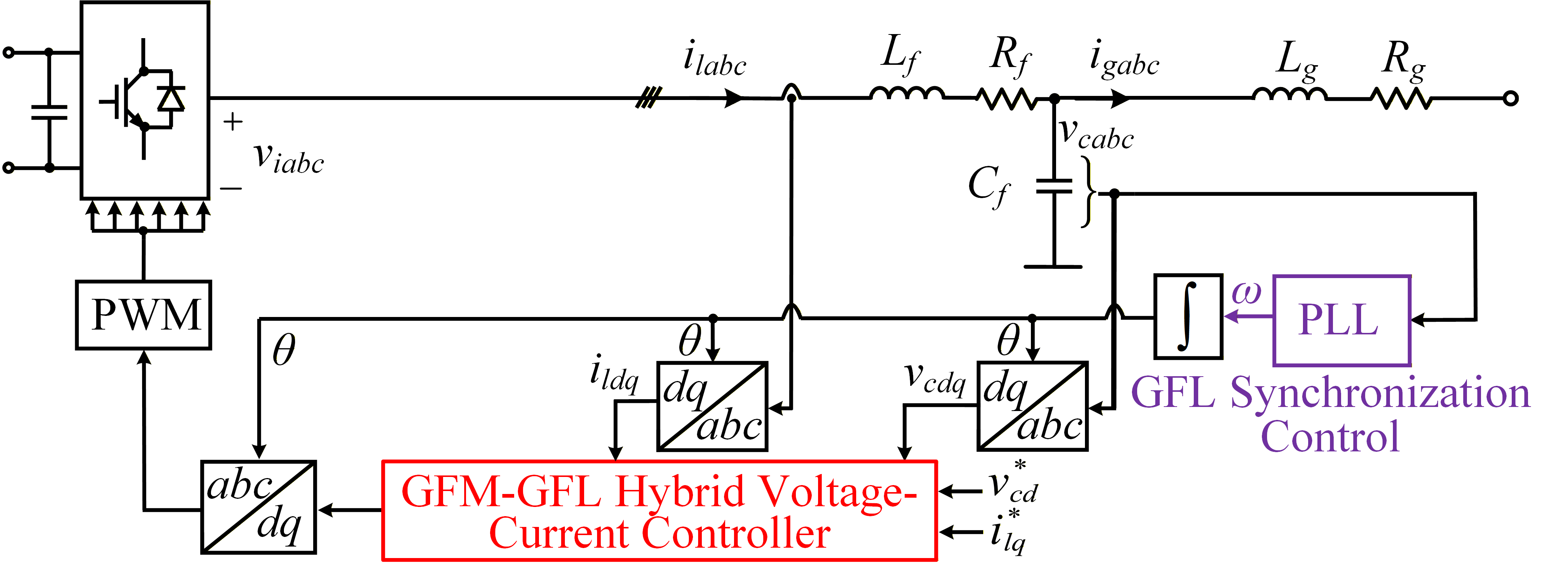}%
\label{fig:vdiqPLL}}
\hfil
\caption{The diagram of four control methods. (a) GFM. (b) GFL. (c) Hybrid magnitude control with $P - \omega$ droop synchronous control. (d) Hybrid magnitude control with PLL synchronous control.}
\label{fig:diagram}
\end{figure*}

\subsection{Small-Signal Stability}\label{Hybrid Voltage-Current Control}
\begin{figure}[htbp]
\centering
\subfloat[]{
 \includegraphics[scale=0.4]{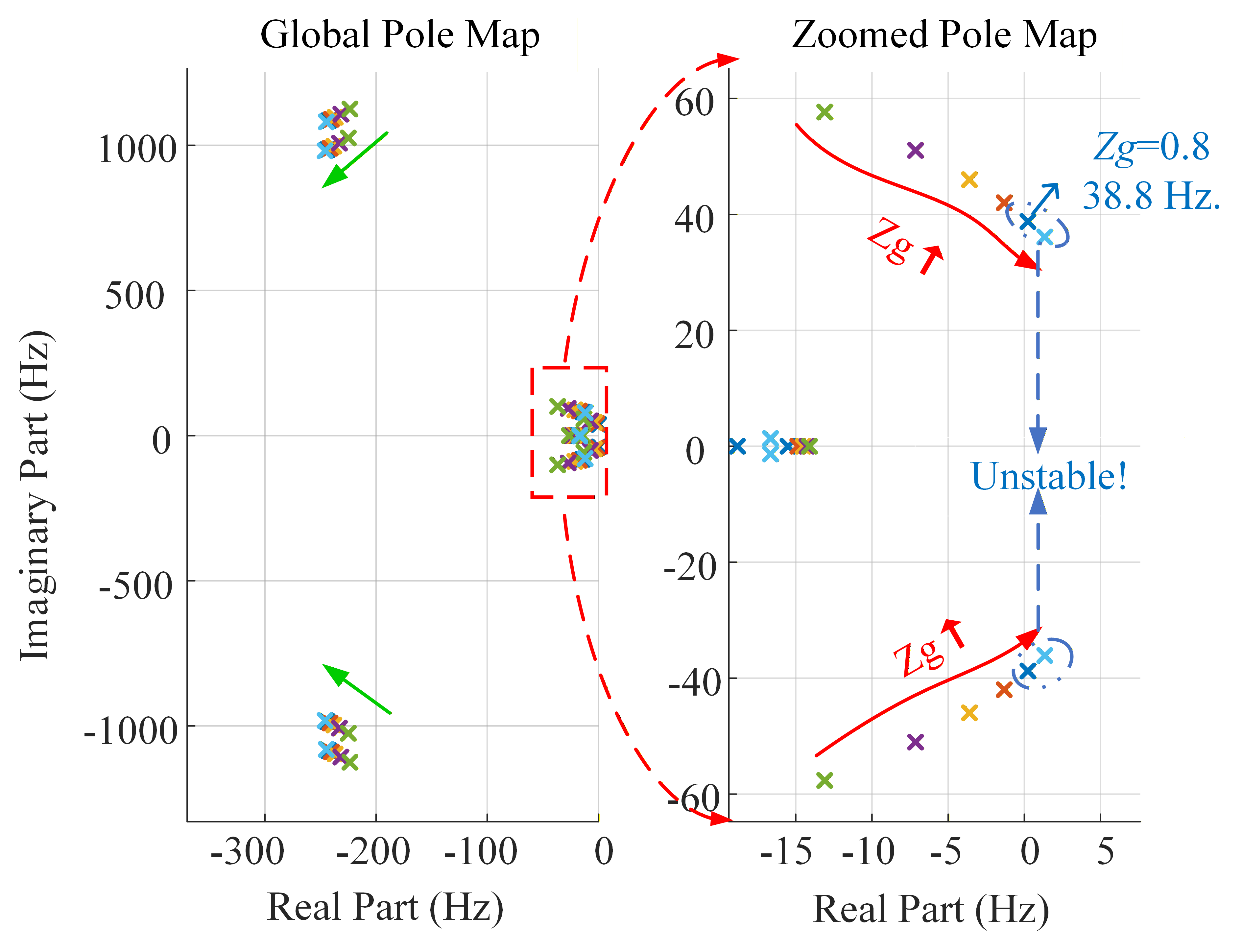}
\label{fig:PoleMap_GFL}
}
\\ [-0.05cm]
\subfloat[]{
 \includegraphics[scale=0.4]{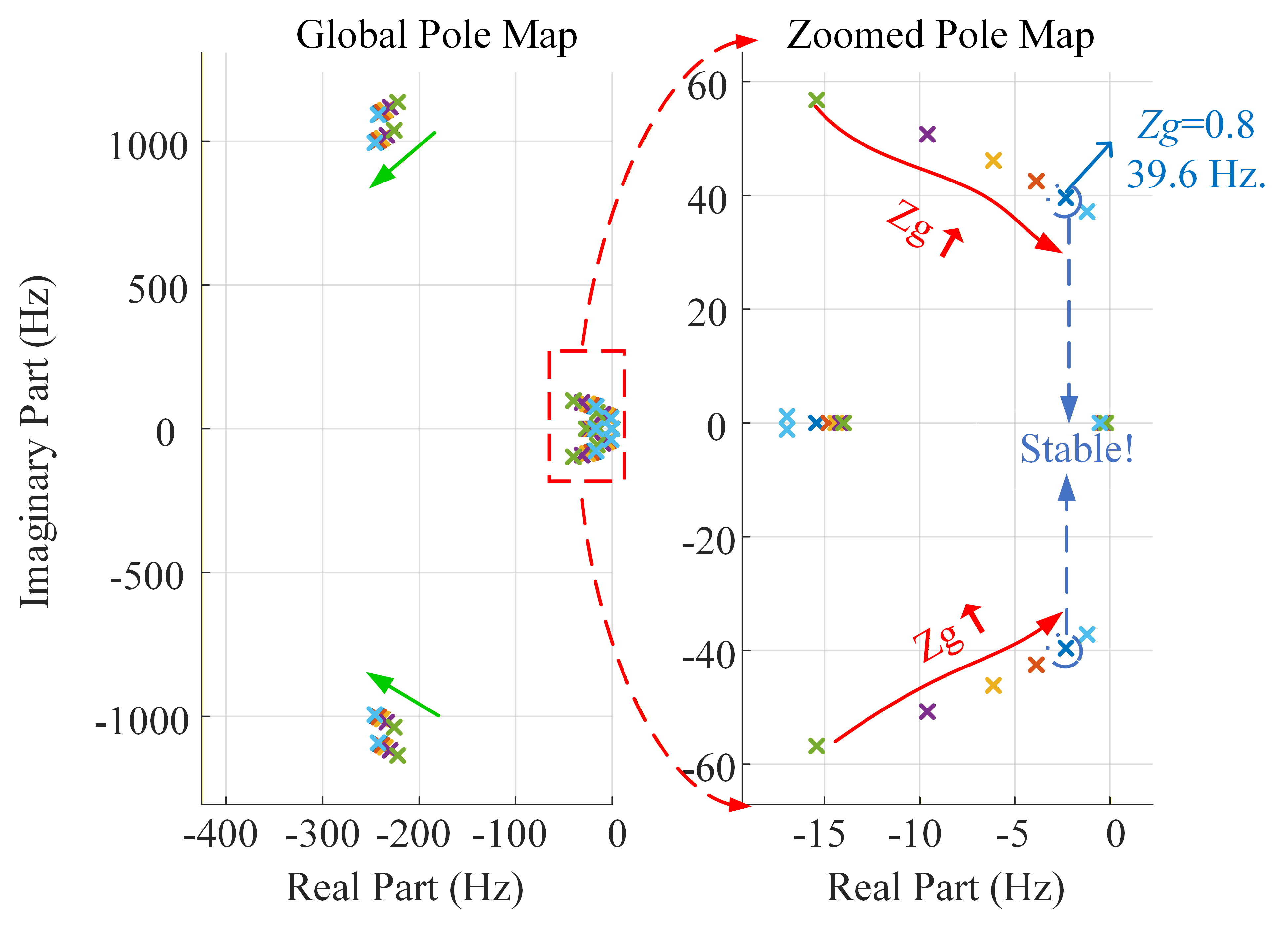}
 \label{fig:PoleMap_AHCGFL}
}
\caption{Pole maps of the two control methods when $Z_g$ varies from 0.2 p.u. to 0.9 p.u..(a) GFL. (b)Hybrid voltage-current control with PLL synchronous control.}
\label{fig:GFLSCR_PoleMap}
\end{figure}

\begin{figure}[htbp]
\centering
\subfloat[]{
    \includegraphics[scale=0.38]{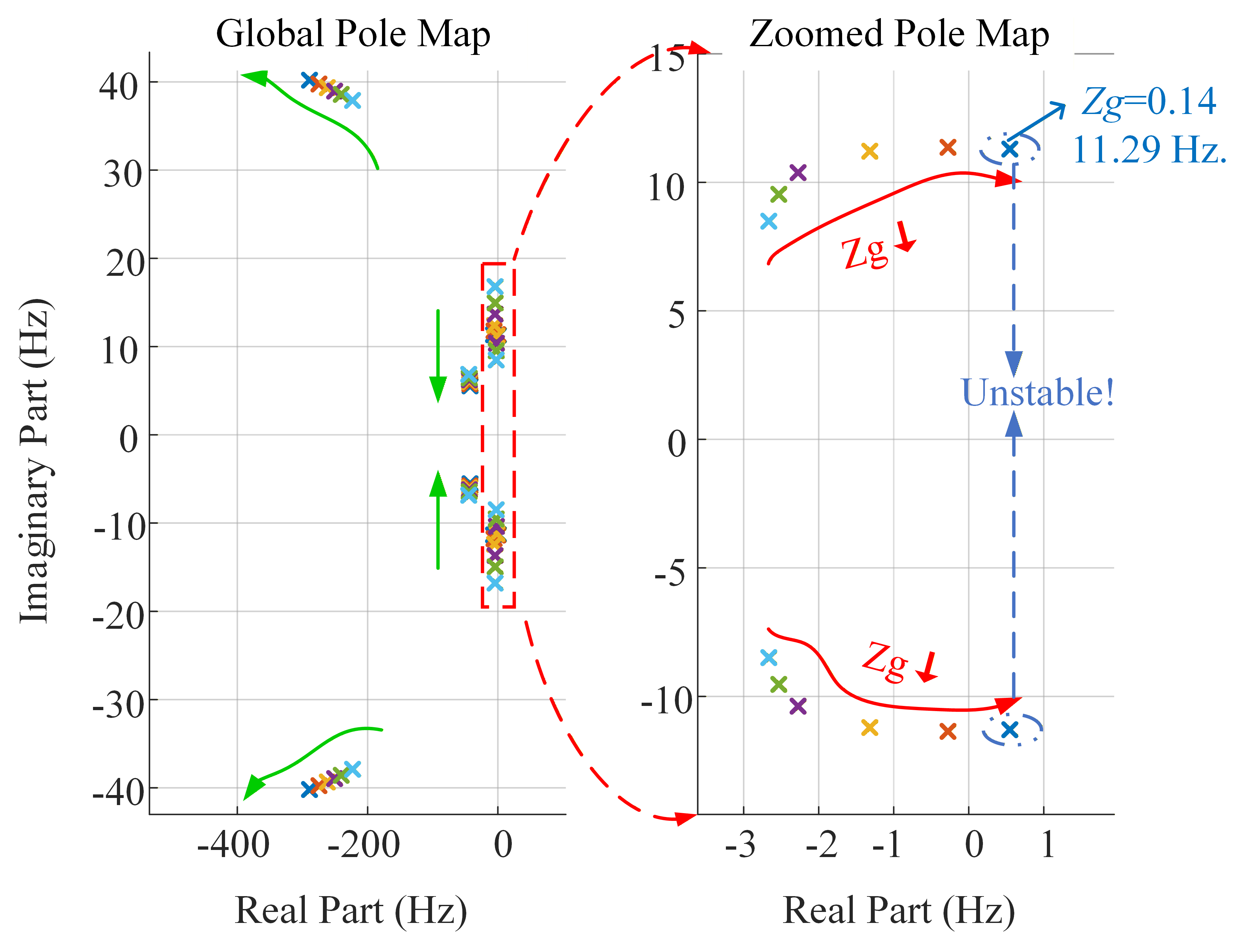}
    \label{fig:PoleMap_GFM}
}
\\ [-0.05cm]
\subfloat[]{
    \includegraphics[scale=0.4]{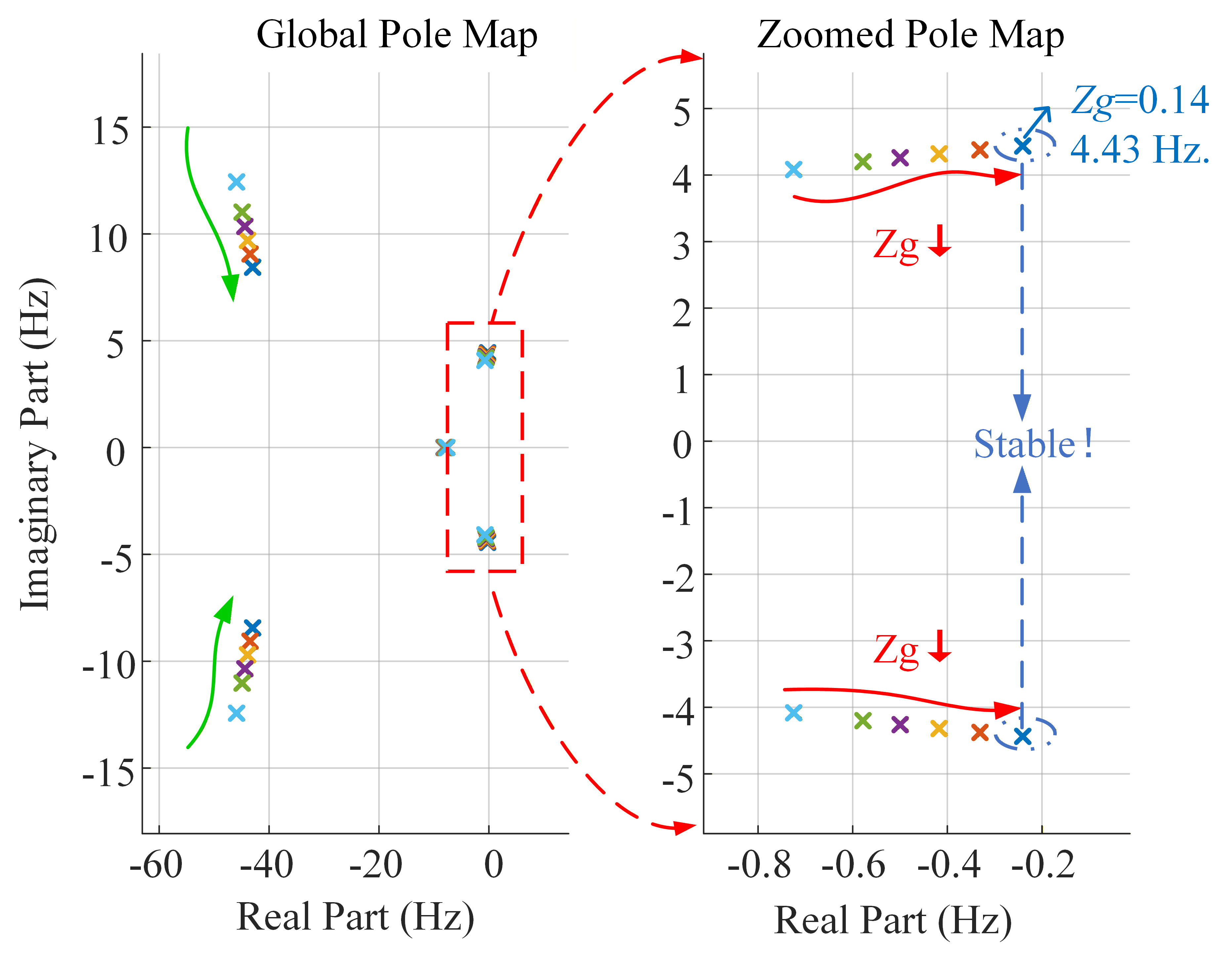}
    \label{fig:PoleMap_AHCGFM}
}
\caption{Pole maps of the two control methods when $Z_g$ varies from 0.2 p.u. to 0.14 p.u..(a) GFM. (b) Hybrid voltage-current control with $P - \omega$ droop synchronous control.}
\label{fig:GFMSCR_PoleMap}
\end{figure}

To demonstrate the superiority of the proposed control strategy, a full-order state-space model is established for the hybrid voltage–current control as well as GFM and GFL control schemes. The model incorporates both the $P-\omega$ droop synchronization and PLL synchronization. The pole maps of the four control architectures under varying line impedance $Z_g$ are illustrated in \figref{fig:GFLSCR_PoleMap} and \figref{fig:GFMSCR_PoleMap}.  
\figref{fig:GFLSCR_PoleMap} shows the results with PLL synchronization. When the line impedance $Z_g$ increases from 0.2 p.u. to 0.9 p.u., the dominant poles of the conventional GFL control move toward the imaginary axis and enter the right-half plane, resulting in oscillatory instability at about 38.8 Hz. However, with the proposed hybrid voltage–current control, the poles remain in the left-half plane and the system remains stable.

\figref{fig:GFMSCR_PoleMap} shows the pole movement with $P-\omega$ droop synchronization. When the line impedance $Z_g$ decreases from 0.2 p.u. to 0.14 p.u., the dominant poles of the conventional GFM control move toward the right-half plane and finally cross the imaginary axis, leading to system instability with an oscillation frequency of about 11.29 Hz. In contrast, under the proposed hybrid voltage–current control, the poles remain in the left-half plane and the system stays stable.
Overall, the proposed hybrid voltage–current control improves the small-signal stability of the system. Under both synchronization schemes, the stability margin is increased and the stable SCR range is expanded.


\section{Simulation Results}
\subsection{Weak-Grid Stability Enhancement Compared with GFL}\label{Weak-Grid Stability Enhancement Compared with GFL}

\begin{figure}[!th]
\centering
\subfloat[]{  \includegraphics[scale=0.39]{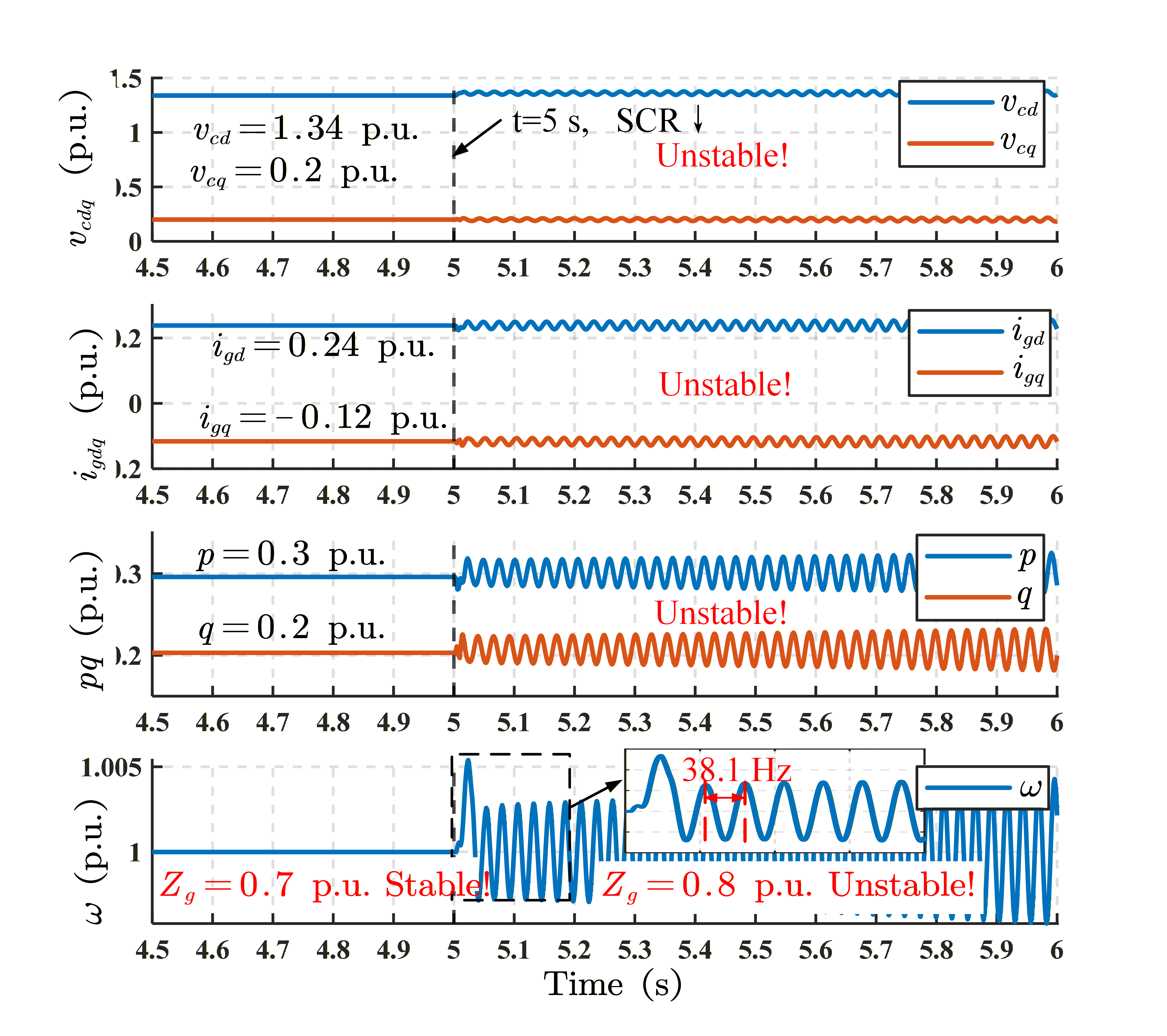}
    \label{fig:waveform_GFL}
}
\\ [-0.05cm]
\subfloat[]{
\includegraphics[scale=0.39]{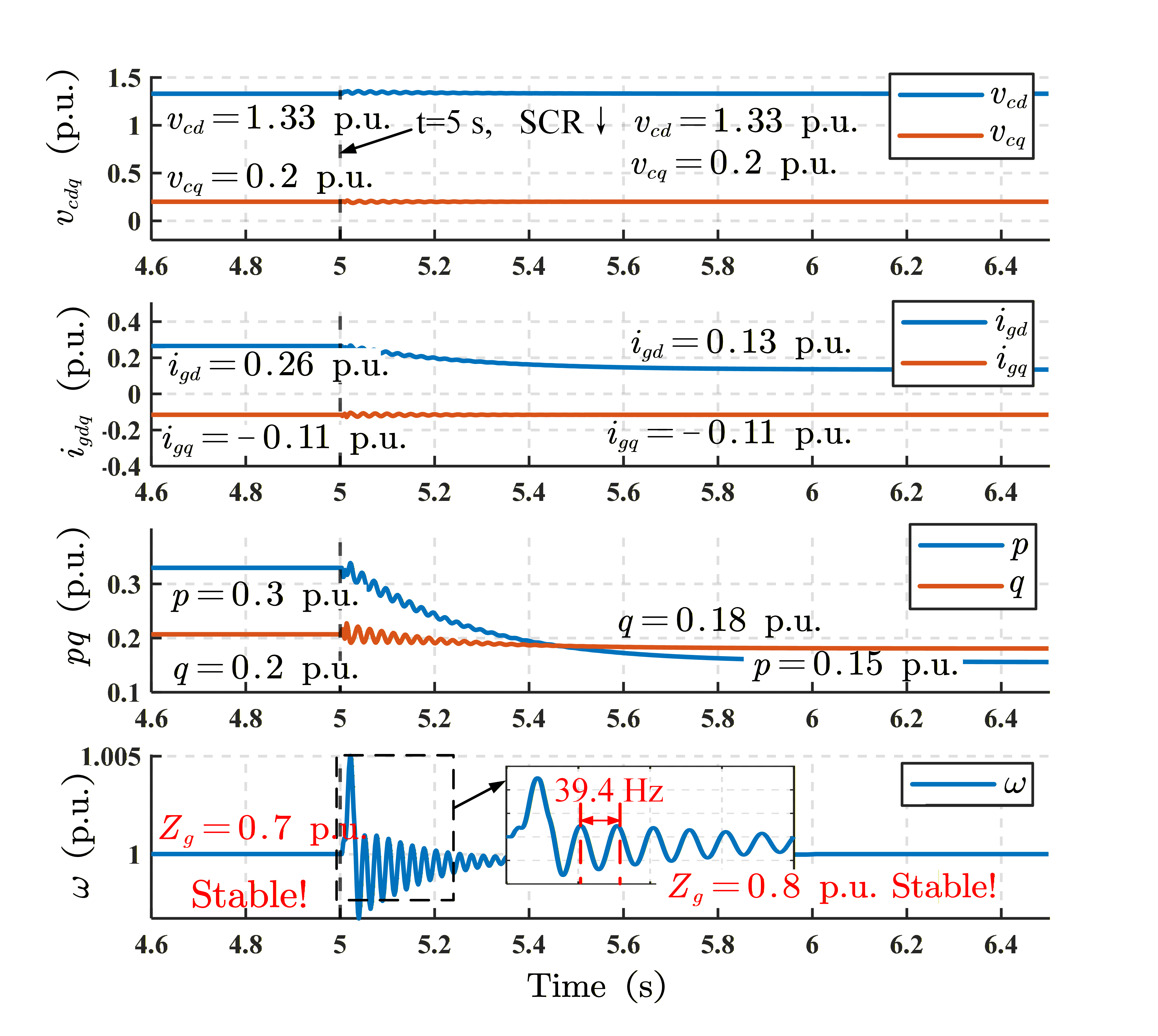}
    \label{fig:waveform_AHCGFL}
}
\caption{The simulation waveforms of the two control methods when SCR decreases from 1.4 to 1.2. (a) GFL. (b) Hybrid voltage-current control with PLL synchronous control.}
\label{fig:GFLSCR_simulationwaveforms}
\end{figure}

Simulations based on MATLAB/Simulink are conducted to compare the stability performance of the hybrid voltage–current control with that of conventional GFM and GFL controls under different grid strength conditions(four control methods are dispicted in \figref{fig:diagram}). The system parameters and control parameters can be found in table \ref{GRID-TIED INVERTER SYSTEM PARAMETERS} and  table \ref{CONTROL PARAMETERS OF FOUR METHOD}.
 \begin{figure}[!thbp]
\centering
\subfloat[]{
    \includegraphics[scale=0.39]{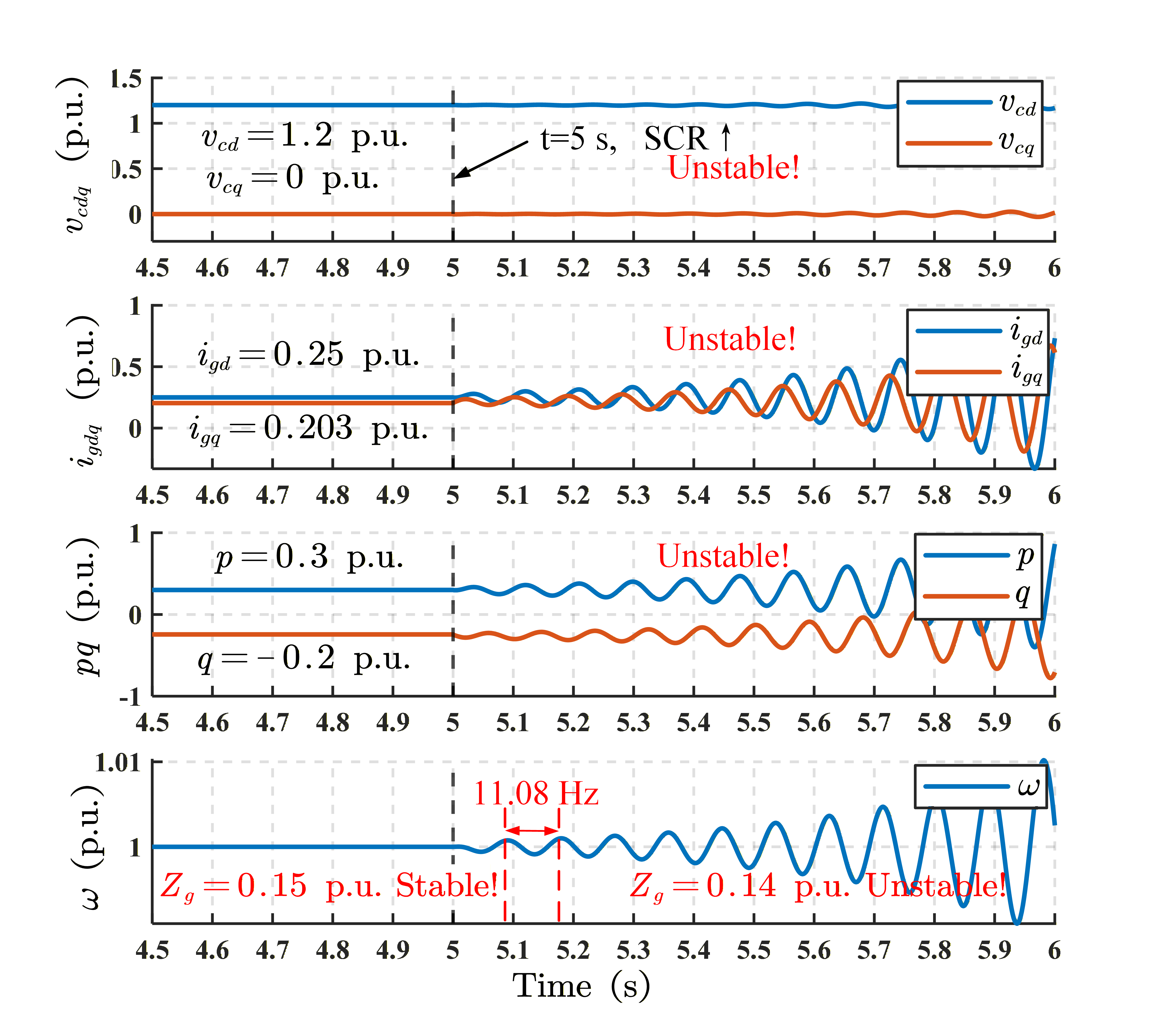}
    \label{fig:waveform_GFM}
}
\\ [-0.05cm]
\subfloat[]{
\includegraphics[scale=0.39]{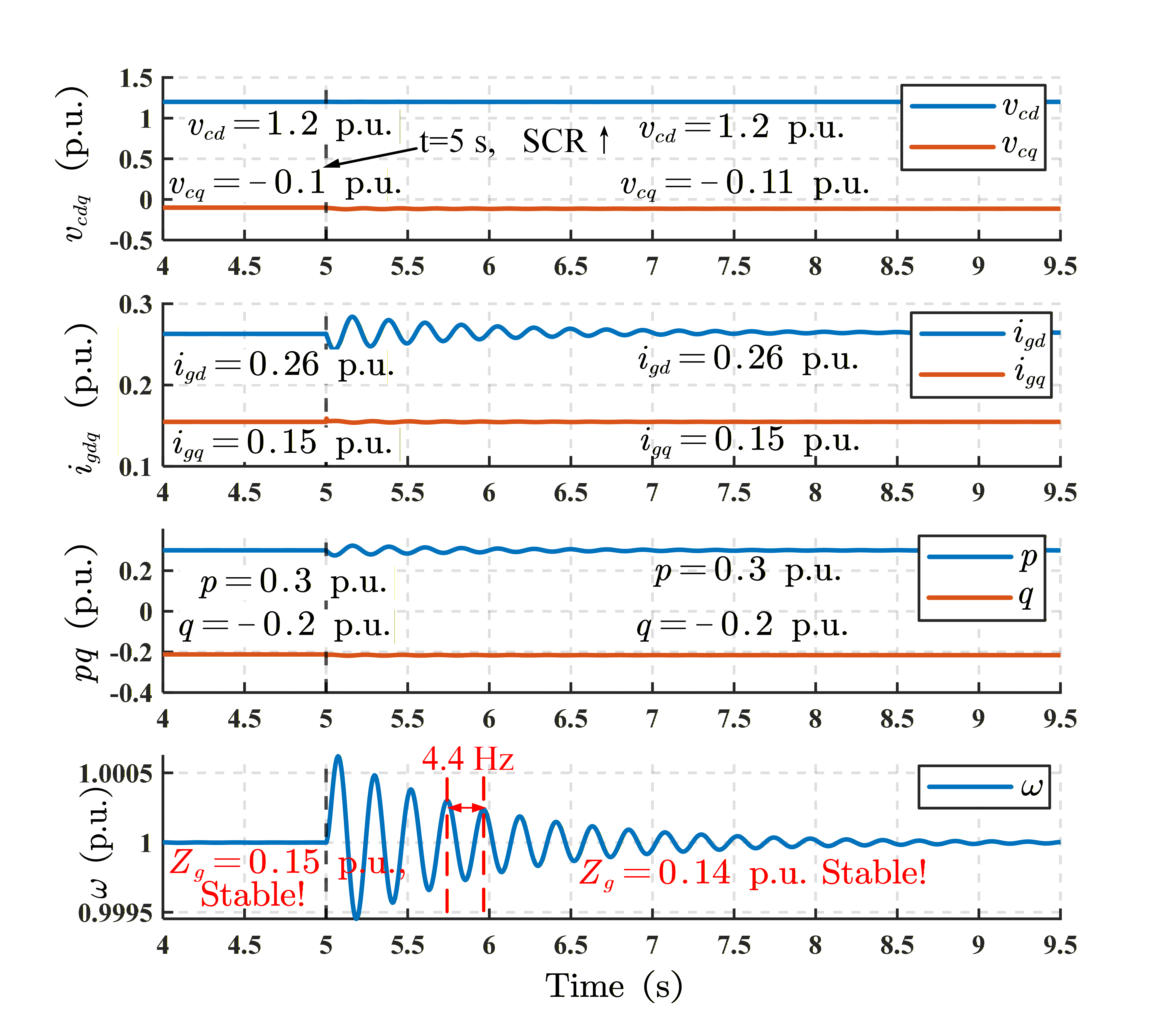}
    \label{fig:waveform_AHCGFM}
}
\caption{The simulation waveforms of the two control methods when SCR decreases from 6 to 7. (a) GFM. (b) Hybrid voltage-current control with $P - \omega$ droop synchronous control.}
\label{fig:GFMSCR_simulationwaveforms}
\end{figure}

To evaluate the weak-grid stability enhancement provided by the proposed hybrid voltage–current control, simulations are carried out under the same steady-state operating point. At $t = 5$ s, the grid strength is suddenly reduced by changing the short-circuit ratio (SCR) from 1.4 to 1.2.

\figref{fig:GFLSCR_simulationwaveforms} (a) shows the dynamic response of the conventional GFL control. Before the disturbance, the system operates stably with $Z_{g}=0.7$ p.u.. When the grid impedance increases to $Z_{g}=0.8$ p.u.(corresponding to a lower SCR), the system becomes unstable with a dominant oscillation frequency of approximately 38.1 Hz. In contrast,\figref{fig:GFLSCR_simulationwaveforms} (b) shows that, under the proposed hybrid voltage–current control, the oscillations(frequency is around 39.4 Hz) are effectively damped and the system converges to a new steady state. These results demonstrate that the proposed control significantly improves the weak-grid stability margin compared with the conventional GFL control, thereby extending the stable operating range of GFL-based systems under weak-grid conditions.

\subsection{Strong-Grid Stability Enhancement Compared with GFM} \label{Strong-Grid Stability Enhancement Compared with GFM}
To further compare the dynamic performance with the grid-forming (GFM) control, time-domain simulations are conducted under the same steady-state operating point. According to \figref{fig:GFMSCR_simulationwaveforms} (a), for the GFM control, the system becomes unstable when the grid impedance decreases from $Z_{g}=0.15$ p.u. to $Z_{g}=0.14$ p.u.. After the disturbance, the system becomes unstable with a dominant oscillation frequency of approximately 11.08 Hz. In contrast, \figref{fig:GFMSCR_simulationwaveforms} (b) shows that, under the proposed hybrid voltage–current control, the system remains stable under the same disturbance and the oscillations(frequency is around 4.4 Hz) gradually decay and the system converges to a new steady state. These results further confirm that the proposed control not only improves the weak-grid stability margin but also broadens the stability region of conventional inverter control strategies.


\section{Experimental Validation} \label{Section:EXPERIMENTAL VALIDATION}
\subsection{Weak-Grid Stability Enhancement Compared with GFL}\label{Weak-Grid Stability Enhancement Compared with GFL}

\begin{figure}[htbp] 
\centering
\includegraphics[width=0.48\textwidth]{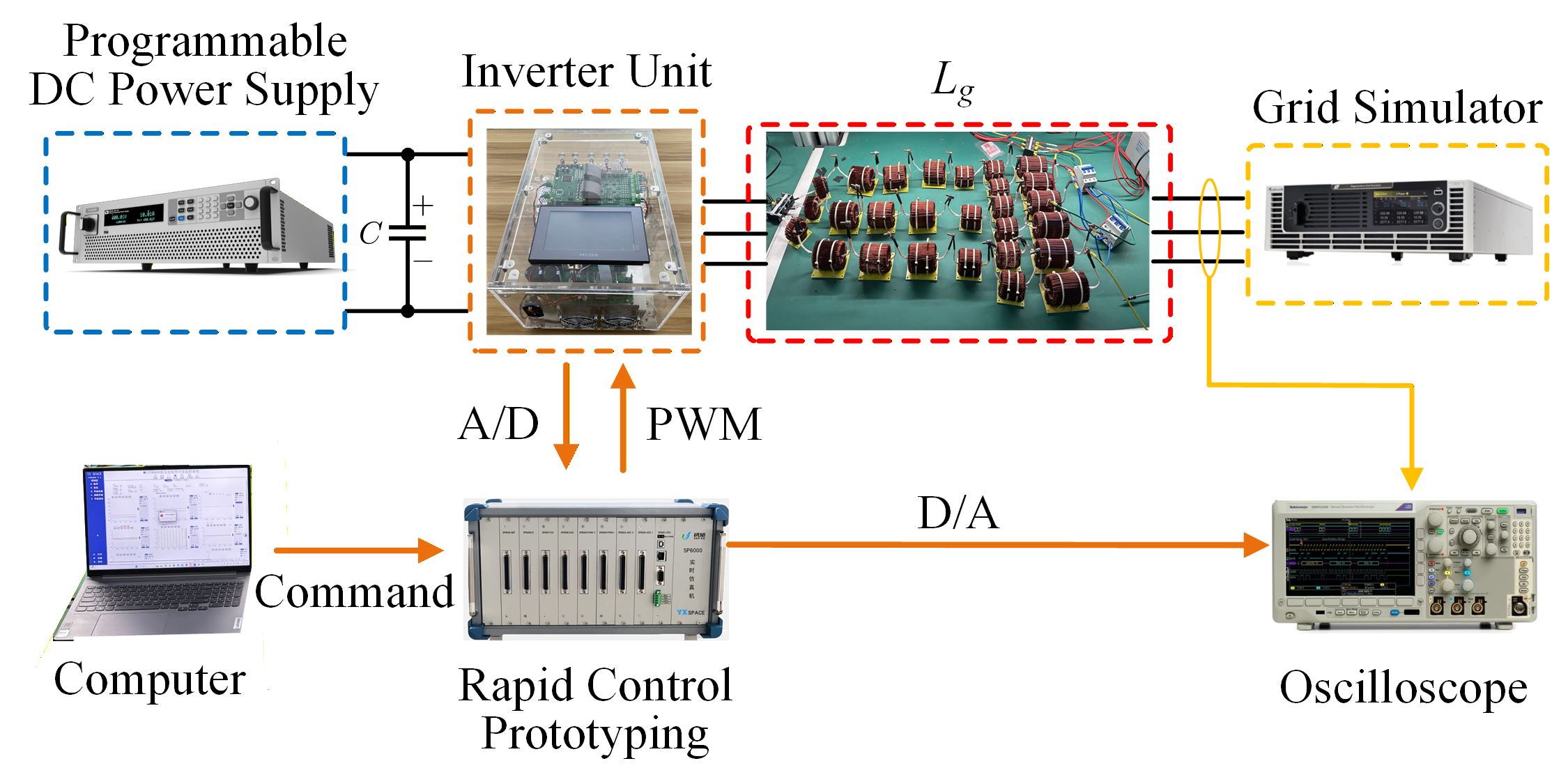} 
\caption{The experimental hardware platform used for verification.}
\label{hardware}
\end{figure}

\begin{figure}[htbp]
\centering
\subfloat[]{
    \includegraphics[scale=0.34]{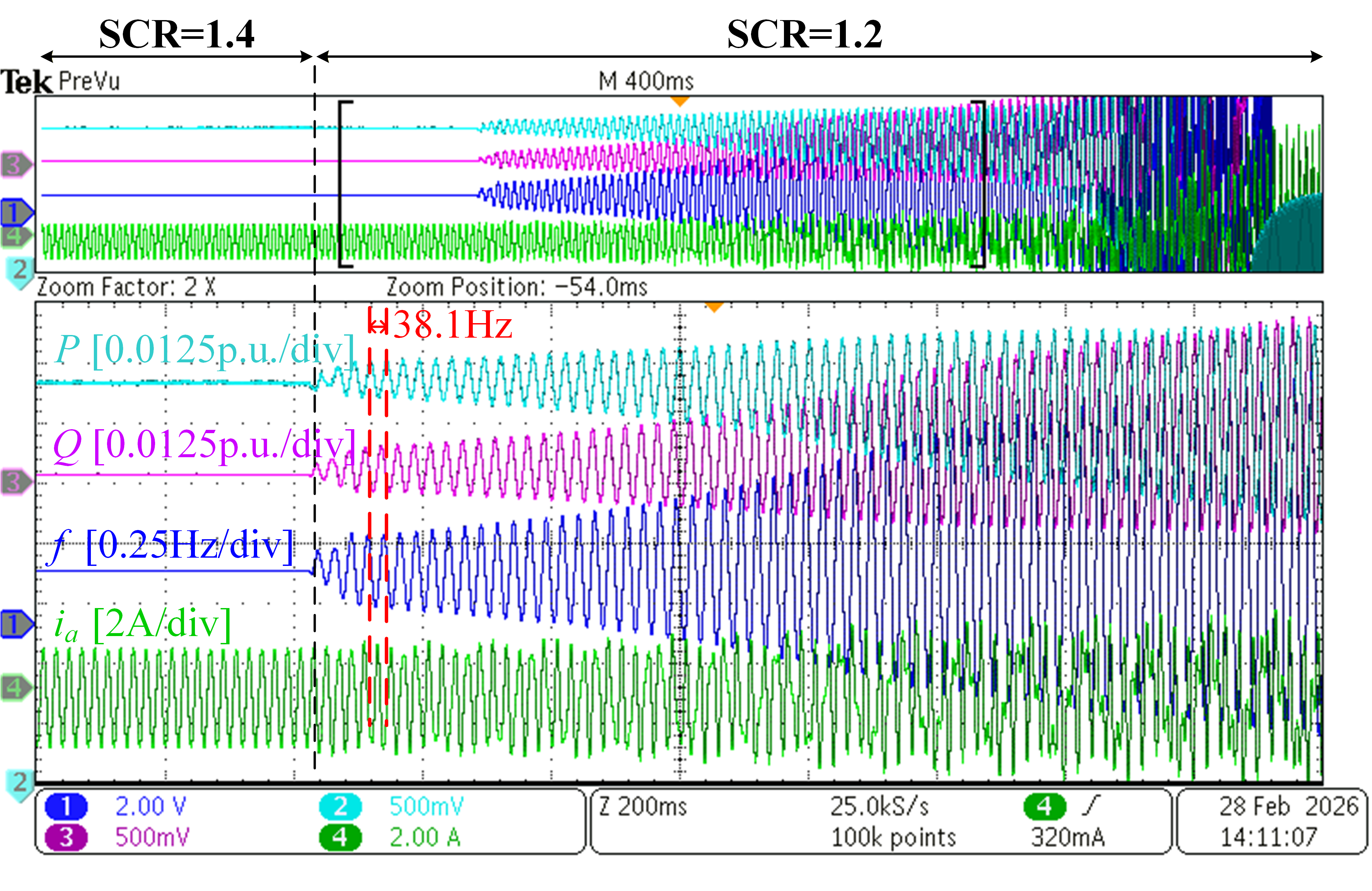}
\label{fig:waveform_GFL_experiment}
}
\\ [-0.005cm]
\subfloat[]{
     \includegraphics[scale=0.34]{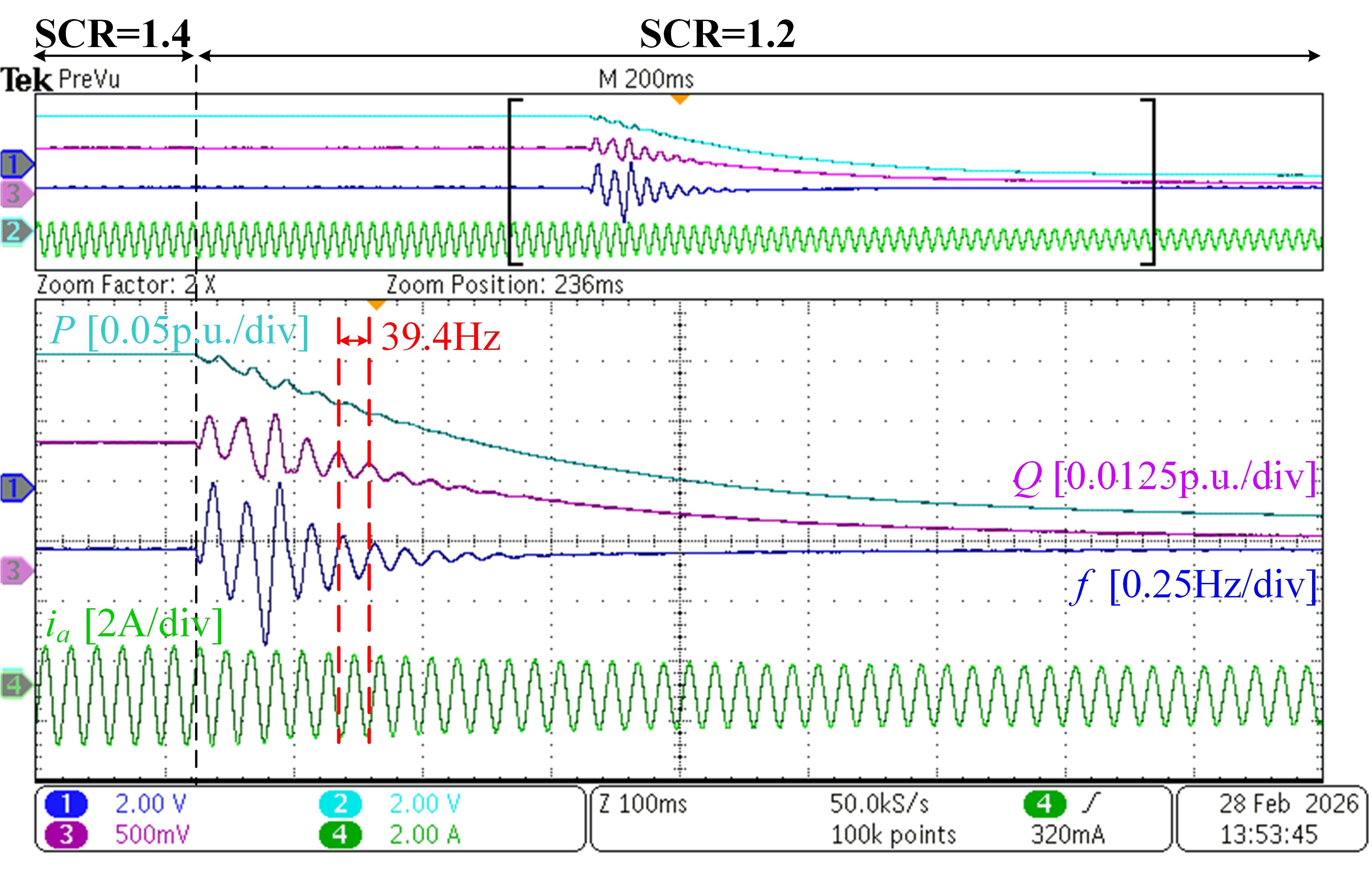}
\label{fig:waveform_AHCGFL_experiment}
}
\caption{The experiment waveforms of the two control methods when SCR decreases from 1.4 to 1.2. (a) GFL. (b) Hybrid voltage-current control with PLL synchronous control.}
\label{fig:GFLSCR_waveforms}
\end{figure}

\figref{hardware} shows the hardware experiment platform. \figref{fig:GFLSCR_waveforms} shows the experimental waveforms under the parameter conditions listed in table \ref{GRID-TIED INVERTER SYSTEM PARAMETERS} and  table \ref{CONTROL PARAMETERS OF FOUR METHOD} when SCR decreases from 1.4 to 1.2. Under the conventional GFL control, a 38.1 Hz oscillation appears and becomes unstable. In contrast, with the proposed hybrid voltage–current control, the system remains stable with a small damped oscillation of about 39.4 Hz. The experimental results show that the proposed hybrid control mitigates the instability issue of conventional GFL control under extremely low SCR conditions caused by the PLL. The proposed control exhibits strong robustness, making it suitable for applications in extremely weak grids, while improving the voltage support capability of the inverter.

\begin{figure}[!h]
\centering
\subfloat[]{
    \includegraphics[scale=0.34]{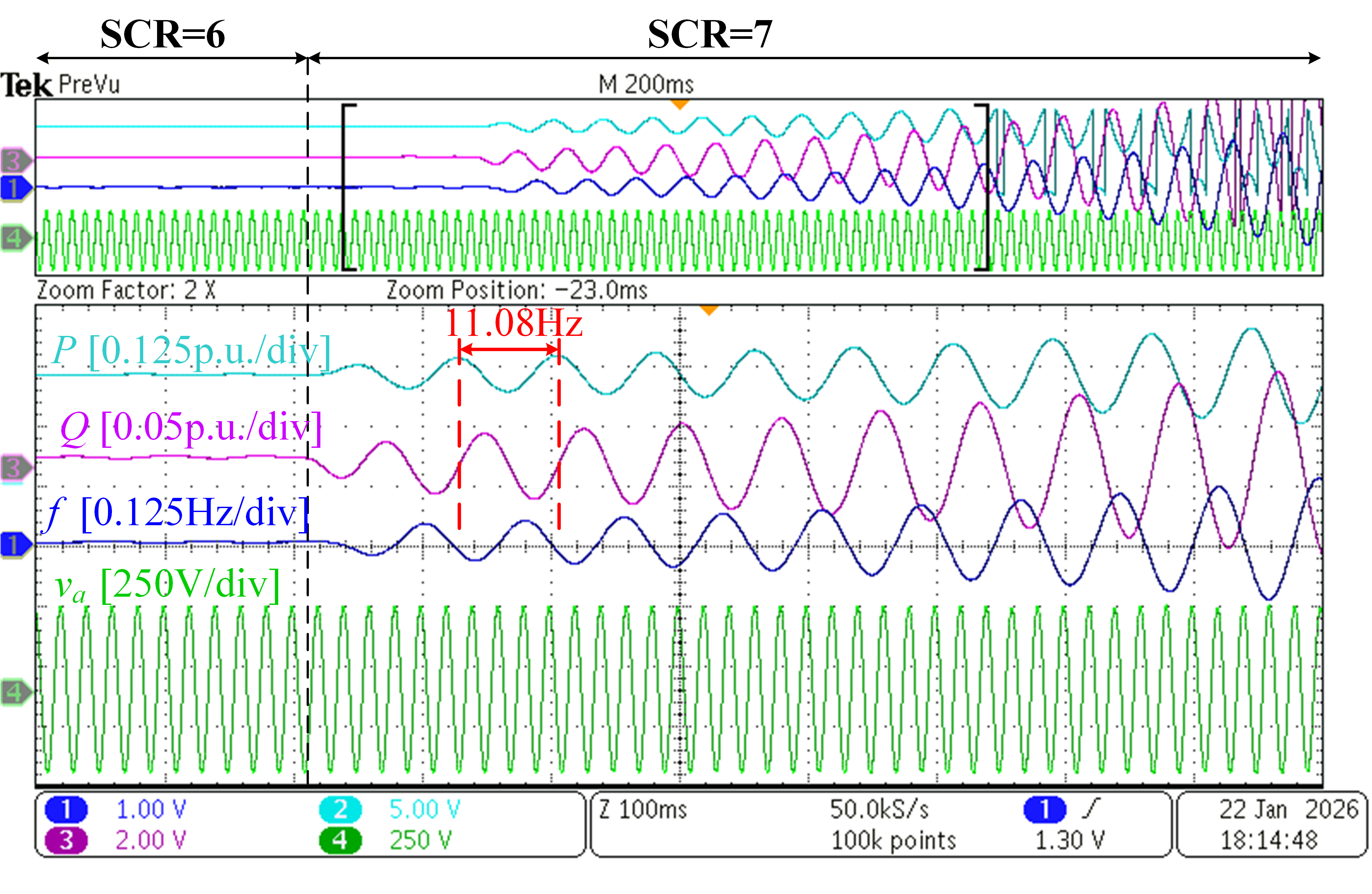}
\label{fig:waveform_GFM_experiment}
}
\\ [-0.005cm]
\subfloat[]{
     \includegraphics[scale=0.34]{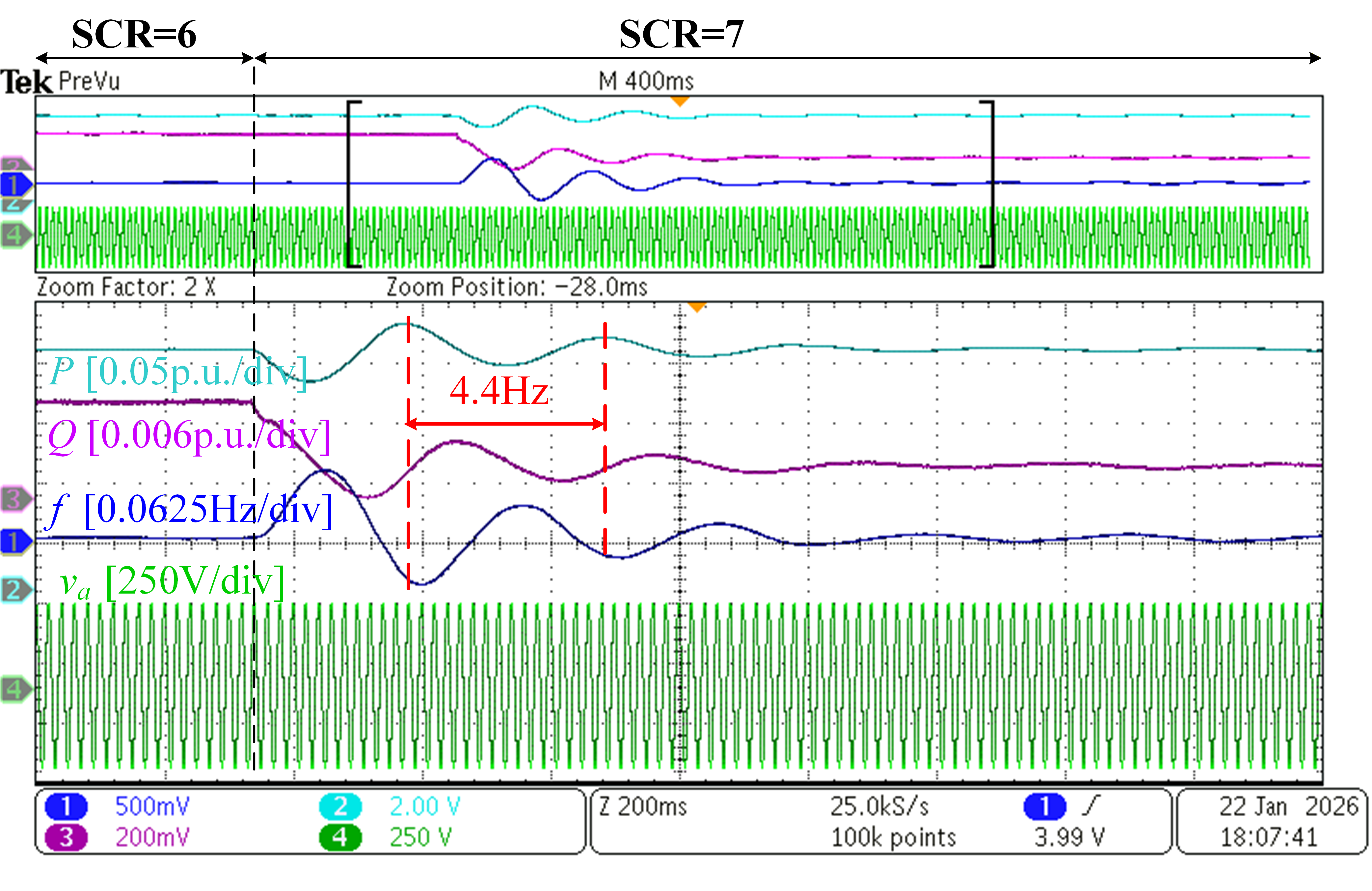}
\label{fig:waveform_AHCGFM_experiment}
}
\caption{The experiment  waveforms of the two control methods when SCR decreases from 6 to 7. (a) GFM. (b) Hybrid voltage-current control with $P - \omega$ droop synchronous control.}
\label{fig:GFMSCR_waveforms}
\end{figure}  

\subsection{Strong-Grid Stability Enhancement Compared with GFM}\label{Strong-Grid Stability Enhancement Compared with GFL}

\figref{fig:GFMSCR_waveforms} shows the experimental waveforms under the same strong-grid parameter conditions when the SCR changes from 6 to 7. Under the conventional GFM control, an oscillation of about 11.08 Hz appears and becomes unstable. However, with the proposed control, the system remains stable with a small damped oscillation of about 4.4 Hz. This comparison indicates that the proposed hybrid control mitigates the low-frequency oscillation issue that conventional GFM control may experience under strong grid conditions. As shown in the waveforms, the proposed control suppresses disturbances and maintains smooth power output during grid condition changes.


\section{Conclusions} \label{Section:Conclusion}

This paper proposes a hybrid voltage–current control method to improve the stability of grid-connected inverters under different grid strength conditions. It integrates the characteristics of conventional GFL and GFM controls. Based on the developed full-order system model, the port characteristics and the small-signal stability of the system are investigated. The experimental results are consistent with the theoretical analysis, which further confirms the effectiveness of the proposed control. The results show that the proposed control mitigates the instability issue of conventional GFL control caused by the PLL under low SCR conditions, while also suppressing the unstable oscillations that may occur in conventional GFM control under strong-grid conditions.


\appendices

\section{System and Control Parameters}

\begin{table}[!htbp]
\centering
\caption{Control Paramters of Inverters}\label{CONTROL PARAMETERS OF FOUR METHOD}
\renewcommand{\arraystretch}{1.6} 
\begin{tabular}{|c|c||c|c|}
\hline
\multicolumn{2}{|c||}{} & \textbf{GFM} & \textbf{Hybrid Control ($\bm{P-\omega}$)} \\ \hline
\multirow{2}{*}{Voltage PI} & $k_{pv}$ & 0.1 & 0.1 \\ \cline{2-4} 
 & $k_{iv}$ & 10 & 10 \\ \hline
\multirow{2}{*}{Current PI} & $k_{pi}$ & 12 & 12 \\ \cline{2-4} 
 & $k_{ii}$ & 5 & 5 \\ \hline
 & $\dfrac{m_p}{2\pi}$ & $2.5 \times 10^{-5}$ & $2.5 \times 10^{-5}$ \\ \cline{2-4} 
\multirow{-2.5}{*}{Synchro.} & $\omega_f$ & $1 \times 2\pi$ & $1 \times 2\pi$ \\ \hline
\hline
\multicolumn{2}{|c||}{} & \textbf{GFL} & \textbf{Hybrid Control (PLL)} \\ \hline
\multirow{2}{*}{Current PI} & $k_{pi}$ & 12 & 12 \\ \cline{2-4} 
 & $k_{ii}$ & 5 & 5 \\ \hline
 & $k_{p,pll}$ & 0.5 & 0.5 \\ \cline{2-4} 
\multirow{-2}{*}{Synchro.} & $k_{i,pll}$ & 12 & 12 \\ \hline
\end{tabular}
\end{table}

\begin{table}[htbp]
\centering
\caption{Parameters of Single-Inverter-Infinite-Bus System}\label{GRID-TIED INVERTER SYSTEM PARAMETERS}
\renewcommand{\arraystretch}{1.3} 
\begin{tabular}{|m{4cm}||c||c|}
    \hline
    \multicolumn{3}{|c|}{\centering \textbf{AC Grid Parameters}} \\ \hline
    \centering Grid Frequency & $f_g$ & 50 Hz \\ \hline
    \centering Grid Voltage & $v_g$ & 200 V \\ \hline
    \multicolumn{3}{|c|}{\centering \textbf{DC-link Parameters}} \\ \hline
    \centering DC-link Capacitor & $C$ & 1.5 mF \\ \hline
    \centering Constant DC-link Voltage & $V_{dc}$ & 500 V \\ \hline
    \centering Rated power & $S_{base}$ & 1500 W \\ \hline
    \multicolumn{3}{|c|}{\centering \textbf{Offset Parameters}} \\ \hline
    \centering Angular Frequency & $\omega^*$ & $2\pi \times f_g$ \\ \hline

    \multicolumn{3}{|c|}{\centering \textbf{$LC$ Filter Parameters}} \\ \hline
    \centering Filtering Inductor & $L_f$ & 3.6 mH \\ \hline
    \centering Inner Resistance of $L_f$ & $R_f$ & 0.08 $\Omega$ \\ \hline
    \centering Filtering Capacitor & $C_f$ & 30 $\mu$F \\ \hline
    \multicolumn{3}{|c|}{\centering \textbf{Grid Line Impedance (Strong-Grid Case)}} \\ \hline
    \centering Line Inductor & $L_g$ & 95 mH \\ \hline
    \centering Inner Resistance of $L_g$ & $R_g$ & 0.33 $\Omega$ \\ \hline
    \multicolumn{3}{|c|}{\centering \textbf{Grid Line Impedance (Weak-Grid Case)}} \\ \hline
    \centering Line Inductor & $L_g$ & 11 mH \\ \hline
    \centering Inner Resistance of $L_g$ & $R_g$ & 0.33 $\Omega$ \\ \hline
\end{tabular}
\end{table}


\ifCLASSOPTIONcaptionsoff
  \newpage
\fi

\bibliographystyle{IEEEtran}
\bibliography{Paper}

\end{document}